\documentclass[aps,prb,reprint,superscriptaddress,showkeys,notitlepage]{revtex4-2}
\usepackage{amssymb}
\usepackage{amsmath}
\usepackage{graphicx}
\usepackage[linktocpage=true,colorlinks=true,pdfborder={0 0 0},linkcolor=blue, citecolor=blue,filecolor=yellow,urlcolor=blue,bookmarks,pdfauthor={},]{hyperref}
\usepackage[dvipsnames]{xcolor}
\usepackage{natbib}
\usepackage[normalem]{ulem}
\newcommand{\Puebla}{Instituto de F{\'i}sica, Benem{\'e}rita Universidad Aut{\'o}noma de Puebla, Apartado Postal J-48, 72570, Puebla, Puebla, M{\'e}xico }
\newcommand{\LV}{Department of Physics \&{} Astronomy, University of Nevada Las Vegas, Las Vegas, Nevada 89154, USA}
\newcommand{\NEXCL}{Nevada Extreme Conditions Laboratory, University of Nevada Las Vegas, Las Vegas, Nevada 89154, USA}

\newcommand{\rev}[1]{\textcolor{red}{#1}} 

\begin{document}

\title{The Isotope Effect and Critical Magnetic Fields of Superconducting YH$_{6}$: A Migdal-Eliashberg Theory Approach}

\author{S. Villa-Cort{\'e}s}
\email{sergio.cortes@unlv.edu, svilla@ifuap.buap.mx}
\affiliation{\Puebla}
\affiliation{\NEXCL}
\author{O. De la Pe{\~n}a-Seaman}
\affiliation{\Puebla}
\author{Keith V. Lawler}
\affiliation{\NEXCL}
\author{Ashkan Salamat}
\affiliation{\NEXCL}
\affiliation{\LV}

\date{\today}

\begin{abstract}

The emergence of near-ambient temperature superconductivity under pressure in the metal hydride systems  has motivated a desire to further understand such remarkable properties, specifically critical magnetic fields. YH$_6$ is suggested to be a departure from conventional superconductivity, due to apparent anomalous behavior.
Using density functional calculations in conjunction with Migdal-Eliashberg theory we show that in YH$_6$ the critical temperature and the isotope effect under pressure, as well as the high critical fields, are consistent with strong-coupling conventional superconductivity; a property anticipated to extend to other related systems. Furthermore, the strong-coupling corrections occur to the expected BCS values for the Ginzburg-Landau parameter ($\kappa_{1}(T)$), London penetration depth ($\lambda_{L}(T)$), electromagnetic coherence length ($\xi(T)$), and the energy gap ($\Delta_{0}$). 

\end{abstract}




\maketitle

Over the last decade, a new class of materials of stoichiometric to hydrogen-rich metal hydrides under pressure has emerged with theoretical predictions being made about their crystal structures, electronic, dynamic, and coupling properties~\cite{cor6,Duan2019,PhysRevB.96.100502,BI2019}.
As a result of those periodic table spanning predictions~\cite{SEMENOK2020100808}, several conventional-superconductor candidates have been proposed with critical temperatures ($T_{c}$) approaching room temperature~\cite{10.1038/srep06968,Liu6990,Wang6463}.
The experimental breakthrough for these materials came with the discovery of phonon-mediated superconductivity in H$_{3}$S, with a maximum $T_{c}$ of $203$~K measured at 155\,GPa~\cite{203,eina,PhysRevB.105.L220502}. Subsequently, high-$T_c$ superconductivity measurements were reported in other compounds such as LaH$_{10}$ with a $250$--$260$~K  $T_c$ at $170$~GPa \cite{La50,PhysRevLett.122.027001}; YH$_{9}$ with $T_c=262$~K at $182$~GPa~\cite{PhysRevLett.126.117003}; YH$_6$ with $T_c=220$~K at approx. $160$~GPa~\cite{Kong2021,Wang_2022}; more recently, CaH$_{6}$ with $T_c=215$~K at $172$~GPa~\cite{PhysRevLett.128.167001}; as well as a reported carbonaceous sulfur hydride with $T_c=287$~K at $267$~GPa~\cite{maxtc,cshret}.
Most of the theoretical works on the superconducting state of these novel metal hydrides have shown that a strong electron-phonon coupling and high-energy hydrogen phonon modes play a key role in the high $T_{c}$ calculated values, concluding that they are phonon-mediated strong-coupling superconductors~\cite{PhysRevB.96.100502,PhysRevB.91.220507,cor4,cor5,PhysRevLett.114.157004,cor1,cor2,Duan2019,CUI20172526,Flores-Livas2016,Szczniak201730,PhysRevB.93.094525,PhysRevB.93.104526,0022-3719-7-15-015,cor6,BI2019,PhysRevB.99.220502}. 

Beyond the $T_{c}$, there are several other important properties of a superconducting material such as the isotope effect coefficient, the upper, lower, and thermodynamic critical magnetic fields, the penetration depth, and the coherence length. 
The first has been crucial in elucidating the mechanism responsible for Cooper-pair formation in conventional superconductors.
While the last gives us the response of the materials to an external magnetic field. 
The lower and upper critical magnetic fields ($H_{c1}$ and $H_{c2}$) are specially interesting, since they give a measure of the Meissner effect and the magnitude of the external magnetic field at which superconductivity is totally suppressed.
Due to experimental difficulties~\cite{Eremets2022}, measurements of the critical magnetic fields of the metal hydrides at high pressures have been done only at temperatures near $T_{c}$.
From that data the slope of $H_{c2}(T)$ is fitted and $H_{c2}(0)$ is extrapolated using the Ginzburg-Landau (GL)~\cite{GL} or the Werthamer–Helfand-Hohenberg (WHH) \cite{WHH} models, giving rise to different values of $H_{c2}(0)$. 
For example, the reported $H_{c2}(0)$ values for YH$_6$, the aim of this work, are 107(157)\,T and 76(102)\,T at 160\,GPa and 200\,GPa, respectively, for GL(WHH)~\cite{Kong2021,adma.202006832}. 
Similar differences are reported in other metal hydrides~\cite{203,La50,Eremets2022}.
Furthermore, these extrapolated values of $H_{c2}(0)$, in conjunction with GL model, Bardeen-Cooper-Schrieffer (BCS) theory~\cite{PhysRev.108.1175}, and empirical relations~\cite{PhysRevB.103.134505,HIRSCH20211353896,doi:10.1063/5.0091404,DOGAN20211353851}, are used to get a complete description of other relevant quantities, like the critical magnetic fields, penetration depth ($\lambda$), coherence length ($\xi$), as well as the GL parameter ($\kappa=\lambda/\xi$).

While the $T_{c}$ of the metal hydrides has been widely studied theoretically within a strong-coupling formalism (Migdal-Eliashberg (ME) theory~\cite{Eliashberg}), there are no reports where these other important properties, like the critical magnetic fields and related lengths, are calculated from first principles within a strong-coupling formalism, where corrections to the empirical and weak-coupling values are expected.
This, in conjunction with the lack of experimental measurements of the critical fields at low temperatures, has generated confusion and some doubts about the nature of the kind of superconductors they belong to~\cite{PhysRevB.103.134505,HIRSCH20211353896,doi:10.1063/5.0091404,DOGAN20211353851}.
For YH$_{6}$, this has even suggested a possible deviation from conventional superconductivity~\cite{adma.202006832}.
The aim of this communication is to show that the critical magnetic fields, the London penetration depth, and the electromagnetic coherence length can be calculated from first principles using density functional theory (DFT) in conjunction with ME theory, showing the conventional nature of YH$_{6}$ and providing a general prescription to describe the superconducting state in the high-$T_{c}$ metal hydrides.

\begin{figure}
\includegraphics[width=8.4cm]{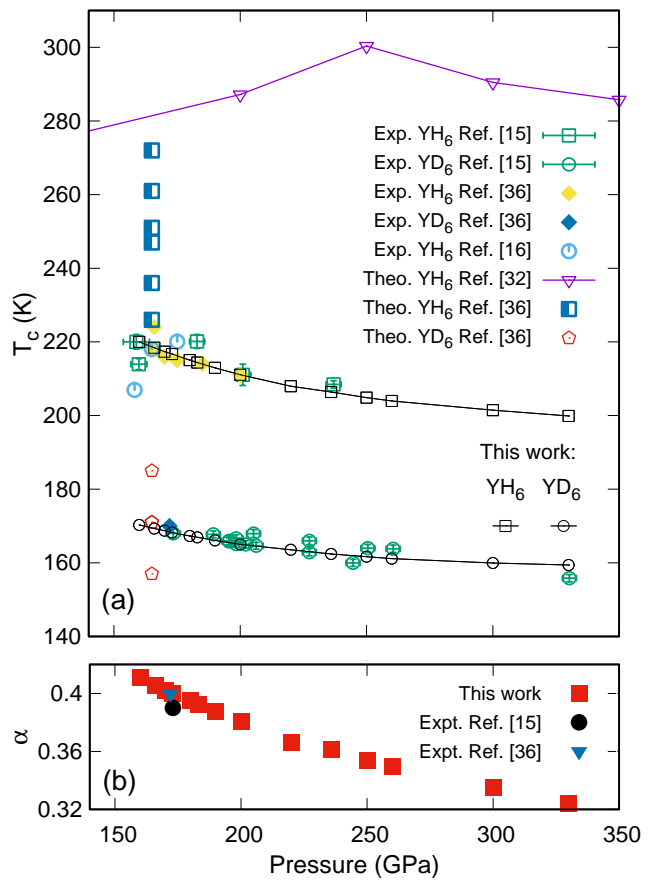}
\caption{\label{fig:TCAL}Calculated $T_{c}(P)$ for both compounds, YH$_{6}$ and YD$_{6}$~(a), and the isotope effect coefficient $\alpha$~(b)  within the ME formalism (Eqs. S18 and S19  in SM~\cite{SM}). For comparison, it is shown the experimental data, as well as previously reported $T_{c}$ calculations.}
\end{figure}

To this end, we start by showing that the behavior of the critical temperature under pressure can be reproduced within the ME formalism, using as input the Eliashgberg function ($\alpha^{2}F\left(\omega\right)$) calculated from first principles in the optimized $lm\bar{3}m$ crystal structure (see Supplemental Material (SM)~\cite{SM}). As a first step, the Linearized Migdal-Eliashberg Equations (LMEE, Eq. S8) are solved at a fixed pressure, where $T_{c}$ is known from experiment. Here, the experimental $T_{c}\,=\,$220~K at 160~GPa \cite{Kong2021} for YH$_{6}$ and $T_{c}\,=\,$165~K at 200~GPa \cite{adma.202006832} for YD$_{6}$ are employed as the initial data. From this solution, the functional derivative of $T_{c}$ with respect to $\alpha^{2}F\left(\omega\right)$ is calculated within the Bergmann and Rainer~\cite{Bergmann1973,LIE1978511} formalism. These functional derivatives allow a link between the observed changes in $\alpha^{2}F\left(\omega\right)$, by a pressure variation (from $P_{0}$ to $P_{i}$), to changes in the critical temperature $\Delta T_{c}(P_{0},P_{i})$ (Eq. S17). Fig.~\ref{fig:TCAL}a shows the behavior of $T_{c}$ under pressure for both, YH$_{6}$ and YD$_{6}$. It can be observed that our calculated $T_{c}$ is in excellent agreement with the experimental values in the whole pressure interval (160\,--\,330~GPa), and even better than the calculations reported previously in the literature, showing that the behavior of $T_{c}(P)$ is driven mainly by the changes in the electron-phonon interaction (Fig.~S4).
The temperature isotope effect coefficient, $\alpha$, is calculated taking into account the changes in the electron-electron and the electron-phonon interaction (Eq. S19), coming from phonon changes due to the isotope mass substitution, within the Rainer and Culetto~\cite{PhysRevB.19.2540, LEAVENS19741329,VILLACORTES2022110451} formalism. Fig. \ref{fig:TCAL}b presents $\alpha$ as a function of pressure, also in excellent agreement with experiment. This first step confirms the conventional nature of superconductivity in YH$_{6}$ and show us that the solutions of the gap function, $\tilde{\Delta}_{n}$, correctly reproduce the superconducting state using the $\alpha^{2}F\left(\omega\right)$ determined from first principles calculations as an input for the LMEEs.

\begin{figure}
\includegraphics[width=8.4cm]{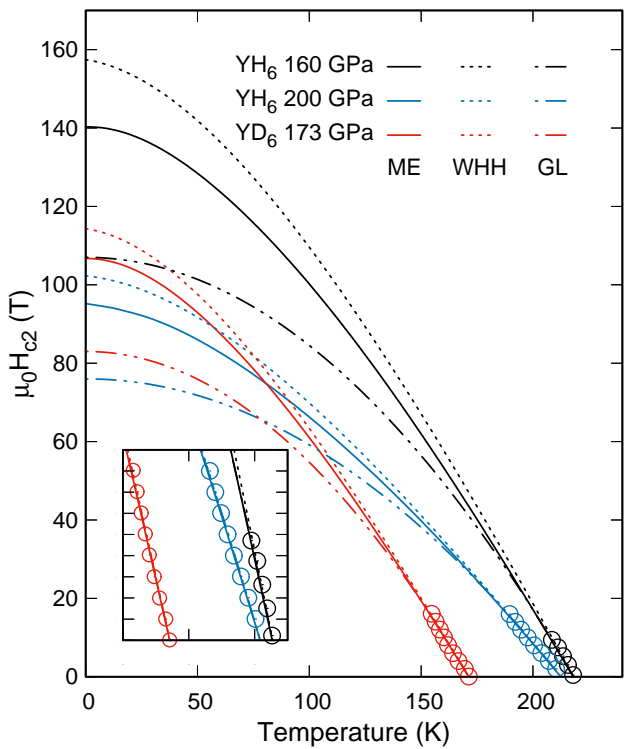}\caption{\label{fig:hc2}Calculated upper critical field for both compounds, YH$_{6}$ and YD$_{6}$, within the Migdal-Eliashberg formalism. For comparison, it is shown the experimental data (symbols)~\cite{Kong2021,adma.202006832} and the extrapolated behavior within the GL and WHH models taken from \cite{Kong2021,adma.202006832}.}
\end{figure}

We now focus on the critical magnetic fields. In the dirty limit for a strong-coupling superconductor, $H_{c2}(T)$ has to be evaluated from the LMEEs (Eqs. S13 and S14) in the presence of a homogeneous magnetic field \cite{Rainer1974} where the pair-breaking parameter $\rho(T)$ and $H_{c2}(T)$ are related by
\begin{equation}
H_{c2}(T)=\rho(T)/eD,\label{eq:hc2}
\end{equation}
\begin{equation}
D=l_{tr}v_{F}/3,\label{eq:diffusion}
\end{equation}
where $D$ is the diffusion constant, $v_{F}$ the Fermi velocity, $l_{tr}$ the mean free path, and $e$ is the electron charge. Figure~\ref{fig:hc2} shows the temperature dependence of $H_{c2}(T)$ calculated within the ME (Eq.~\ref{eq:hc2}, solid lines), GL (dashed lines), and WHH (doted lines) formalism for YH$_{6}$ at 160 and 200~GPa, and YD$_{6}$ at 173~GPa. These pressures were selected to make a direct comparison with available experimental data (symbols)~\cite{Kong2021,adma.202006832}. $v_{F}$ was calculated from the dispersion of the electronic band structure (see Table~\ref{tab:tcero}), and $l_{tr}=1.655(1.645)$\,\AA{} for YH$_6$(YD$_6$) was fitted using Eq.~\ref{eq:diffusion} to get the experimental $H_{c2}$ data close to $T_{c}$. 
Near $T_{c}$, the calculated values of $H_{c2}(T)$ are similar for the three models, and in very good agreement with experiment. As the temperature starts to decrease, the difference between the models' results increases,  with the ME results being in between the higher WHH and lower GL values.
The largest differences between ME and the GL and WHH models ($\Delta{H_{c2}}$) are at $T=0$~K, with values that go from $\Delta{H_{c2}}(\mbox{YH$_6$})=7(18)$~T at 200~GPa to as high as $\Delta{H_{c2}}(\mbox{YH$_6$})=17(33)$~T at 160~GPa, in respect to WHH(GL). Although the GL theory is applicable to practically all superconductors, it is a phenomenological theory restricted to temperatures close to $T_{c}$. Therefore, GL theory is not expected to give accurate results at lower temperatures, giving rise to the huge differences at T=0~K between itself and the other theories. The WHH formalism is based on weak-coupling BCS theory, and thus should be valid for the whole temperature range; whereas the ME formalism covers weak- and strong-coupling. Thus, the ME results show a distinct strong-coupling correction to the WHH results.

\begin{figure*}
\includegraphics[width=17cm]{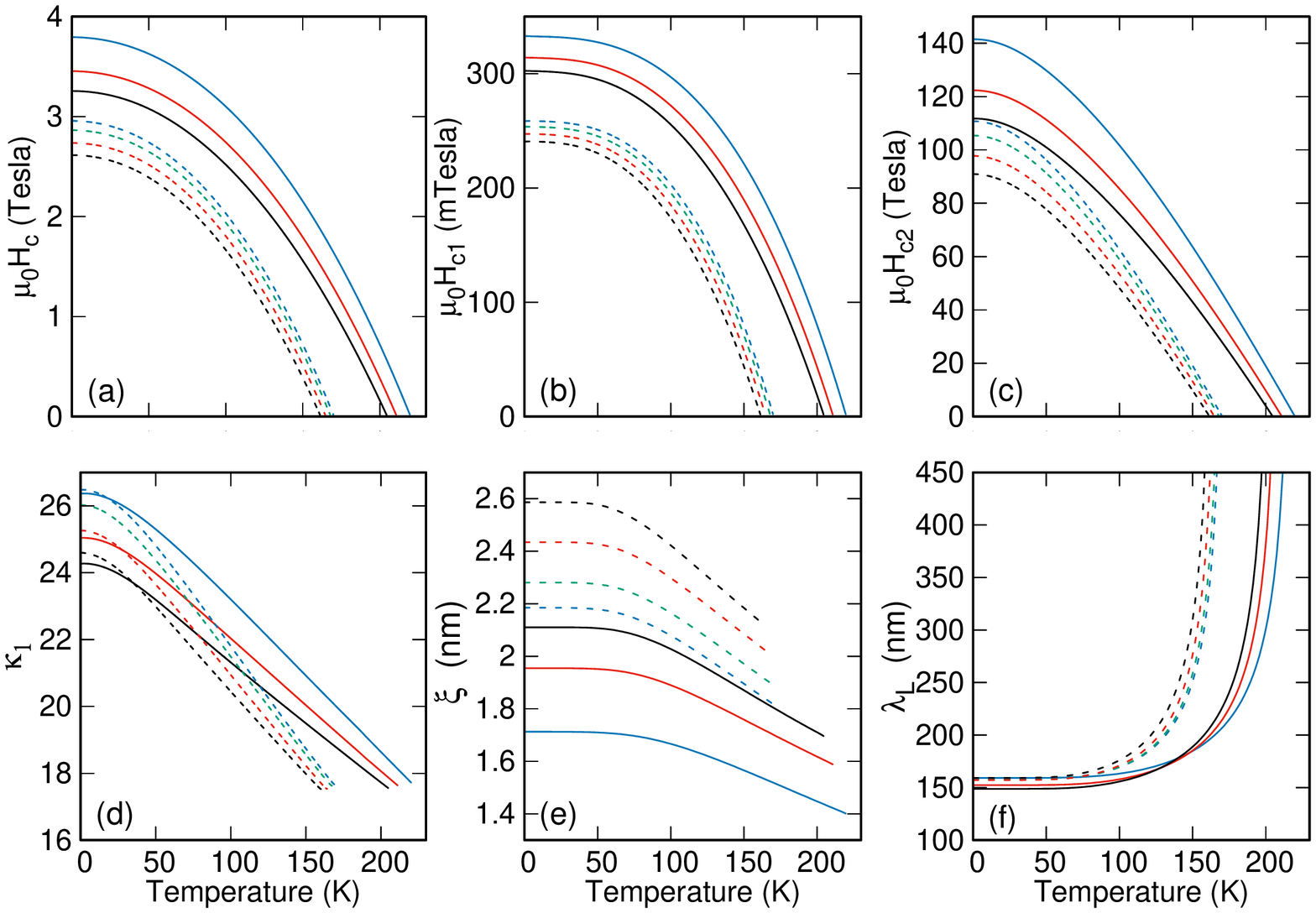}\caption{\label{fig:panel} Calculated thermodynamic (a), lower (b), and upper (c) critical magnetic fields within the Migdal-Eliashberg formalism for YH$_6$ (solid lines) and YD$_6$(dashed lines) at 160 (blue), 173 (green), 200 (red)\rev{,} and 250 (black) GPa. The parameter $\kappa_{1}(T) = (1/\sqrt{2})H_{c2}(T)/H_{c}(T)$ (d) is used in conjunction with the relation $H_{c1}H_{c2} = H_{c}^{2}\text{ln}(\kappa)$ to get the lower critical magnetic field (b). Also shown are the electromagnetic coherence length (e) and the London penetration depth (f) in the clean limit.}
\end{figure*}

To calculate the thermodynamic critical field, $H_{c}$, of an isotropic strong-coupling superconductor, the Non-linear MEEs (NLMEE) have to be solved (Eqs. S5 and S6). With the knowledge of $\tilde{\Delta}_{n}$ and $\tilde{\omega}_{n}$, the difference in free energy between the normal and the superconducting states of the metal, $\Delta F(T)=F_{n}-F_{s}$, can be calculated directly. By definition, $H_{c}$ is given by the relation $H_{c}(T)=[8\pi (\Delta F(T))]^{1/2}$. Then, from the calculated fields, the GL parameter and the lower critical field can be evaluated from the relations $\kappa_{1} = (1/\sqrt{2})H_{c2}/H_{c}$ and $H_{c1}H_{c2} = H_{c}^{2}\text{ln}(\kappa)$, respectively.
Expanding Figure \ref{fig:hc2} to higher pressures to span the whole pressure range studied, Figure \ref{fig:panel} shows the calculated behavior of $H_{c}(T)$, $H_{c1}(T)$, and $H_{c2}(T)$ for YH$_{6}$ (solid lines) and YD$_{6}$ (dashed lines) at 160, 200, and 250~GPa. As a function of pressure, there is a steady shift of the critical fields to lower values as the pressure is incremented. In particular, at $T$=0~K there is a considerable suppression in $H_{c2}$ from 141.5(110.8)~T to 111.8(91.0)~T for YH$_{6}$(YD$_{6}$), at 160 and 250~GPa respectively (see Table~\ref{tab:tcero}). As it can be seen, the three fields show an isotopic shift to lower values due to the replacement of hydrogen by deuterium. 

The strong-coupling behavior of the calculated $H_{c2}(T)$ and $H_{c}(T)$ by the ME-model is confirmed by the deviation function, $D(t)$ (Eqs. S27 and S28), which shows the standard behavior (positive values) for strong-coupling materials, in contrast to the intermediate-coupling behavior (change in sign) that is observed in the WHH model (Fig. S6). 
Fig. \ref{fig:panel}d shows the parameter $\kappa_{1}(T)$, which has values between approximately 24 and 27 for both compounds at $T$=0~K (Table~\ref{tab:tcero}), while at $T_{c}$ (where $\kappa_{1}$ is similar to the GL $\kappa$ \cite{PhysicsPhysique}) this parameter has its minimum values and vary from 17.5 to 17.7. The ratio $\kappa_{1}(0)/\kappa_{1}(T_{c})$ give us an estimation of the strong-coupling correction to BCS values. While in the weak-coupling formalism there is a universal ratio $\kappa_{1}(0)/\kappa_{1}(T_{c})$=1.12 \cite{Rainer1974}, the values for YH$_6$(YD$_6$) vary from 1.49(1.5) to 1.38(1.4) at 160 and 250~GPa, respectively. Such enhanced values clearly show that these systems are within the strong coupling regime across this noted pressure range, and the slight decrease with respect to pressure shows the tendency towards a less strong-coupling (intermediate) regime, as expected from the general trend under pressure of the coupling parameter (Fig. S4).

Despite expressions which are valid within Migdal-Eliashberg theory being derived several years ago~\cite{PhysRev.156.470,PhysRev.156.487}, there are very few strong-coupling superconductor systems with reported numerical results of their electromagnetic properties. Here we are interested in the magnetic field penetration depth (in the London limit) $\lambda_{L}(T)$, and the electromagnetic coherence length $\xi(T)$. Both quantities are studied in the clean limit, due to the very large energy scale associated with the phonon frequencies and superconducting gap for the metal hydrides, as was previously suggested for H$_{3}$S by Nicol and Carbotte~\cite{PhysRevB.91.220507}. The London-limit penetration depth, which applies when $\lambda_{L}(T)\gg\xi(0)$, is given by
\begin{equation}
\lambda_{L}(T)=\left[\frac{4}{3}\pi N(0) e^{2} v_{F}^{2} T \mu_{0}\sum_{n=1}^{\infty}  \frac{\tilde{\Delta}_{n}^{2}}{\left(\tilde{\omega}_{n}^{2} + \tilde{\Delta}_{n}^{2}\right)^{3/2}}\right]^{-1/2},\label{eq:london}
\end{equation}
where $\mu_{0}$ is the permeability and $N(0)$ is the single spin density of electronic states at the Fermi energy. The electromagnetic coherence length, which describes the nonlocality in the electromagnetic response of a superconductor, is given by 
\begin{equation}
\xi(T)=\frac{v_{F}\hslash}{2} \frac{\left[\sum_{n=1}^{\infty}  \frac{\tilde{\Delta}_{n}^{2}}{\left(\tilde{\omega}_{n}^{2} + \tilde{\Delta}_{n}^{2}\right)^{3/2}}\right]}{\left[\sum_{n=1}^{\infty}  \frac{\tilde{\Delta}_{n}^{2}}{\tilde{\omega}_{n}^{2} + \tilde{\Delta}_{n}^{2}}\right]}.\label{eq:cohe}
\end{equation}

The solutions of the NLMEEs (Eqs. S5 and S6), $\tilde{\Delta}$ and $\tilde{\omega}$, are required to get both $\lambda_{L}(T)$ and $\xi(T)$.
Numerical results for the temperature variation of $\xi(T)$ are given in Fig. \ref{fig:panel}e. In superconductors, $\xi(T)$ is the range of the perturbation of the current density caused by an applied electromagnetic field, which is different from the GL coherence length $\xi$ that describes the perturbation of the superconducting pair density. In the weak-coupling theory, both quantities are related at $T=0$~K through the relation $\xi=0.739\xi(0)$. From these strong-coupling formalism calculations,  $\xi(0)$ is found to be 1.71(2.18)~nm and 2.11(2.58)~nm for YH$_{6}$(YD$_{6}$) at 160 and 250~GPa, respectively.
Using the GL relation, $\xi=\sqrt{\frac{\phi_{0}}{2\pi H_{c2}}}$, the GL coherence length $\xi$ is found to be 1.53(1.72)~nm and 1.73(1.94)~nm for  YH$_{6}$(YD$_{6}$), at the same applied pressures which are in agreement with GL experimental estimations of 1.4--1.8~nm at 160~GPa for YH$_6$~\cite{Kong2021,adma.202006832}.
These values yield ratios $\xi/\xi(0)$ 0.895(0.787) and 0.82(0.75) for YH$_{6}$(YD$_{6}$) at 160 and 250~GPa, respectively, which clearly deviate from the weak-coupling limit of 0.739. As the temperature increases, $\xi(T)$ falls quickly near $T_{c}$, where the ratio $\xi(T_{c})/\xi(0)$ drops to a value of about 0.82(0.83) for YH$_{6}$(YD$_{6}$) at 160 GPa, which is larger than the BCS value of 0.752~\cite{PhysRevB.18.6057}. 

Fig.~\ref{fig:panel}f shows the temperature dependence of $\lambda_{L}(T)$. 
$\lambda_{L}(0)$ is found to be approximately 159~nm for both, YH$_{6}$ and YD$_{6}$ at 160~GPa, which is close to the reported BCS values of 164 and 147~nm for H$_{3}$S and LaH$_{10}$, respectively~\cite{DOGAN20211353851}. The strong-coupling deviation function of $[\lambda_{L}(0)/\lambda_{L}(T)]^{1/2}$ with respect to the two fluid model for YH$_{6}$ is shown in Fig. S6. The deviation function (Eq. S29), gives a minimum value of -0.05~GPa, which is in contrast to the -0.22~GPa result from  BCS weak-coupling theory~\cite{RevModPhys.62.1027}. 
By means of Pad\'{e} approximants~\cite{pade}, the energy gap, $\Delta_{0}$, can be found from an analytic continuation to the real axis of $\tilde{\Delta}_{n}$ (computed from the NLMEEs). Here, $\Delta_{0}$ is 46.68(37.6)~meV and 37.7(30.9)~meV for YH$_{6}$(YD$_{6}$) at 160 and 250~GPa, respectively. From these values, the BCS ratios $2\Delta_{0}/k_{B}T_{c}$ are 4.94(5.13) and 4.50(4.74) which shows a strong-coupling correction to the 3.52 BCS value.

\begin{table*}
\caption{\label{tab:tcero}Zero-temperature calculated parameters of the superconducting state at selected pressures for YH$_{6}$ and YD$_{6}$ using the solutions of the NLMEEs and LMEEs, in conjunction with Eqs. \ref{eq:hc2}, \ref{eq:london} and \ref{eq:cohe} for the upper critical field, London penetration depth ($\lambda_{L}$), and coherence length ($\xi$), respectively. The critical fields are in units of Tesla, the Fermi velocity ($v_{F}$) in $\times10^{5}m/s$ (Eq. S26), $\lambda_{L}$ and $\xi$ in nm.}
\begin{ruledtabular}
\begin{tabular}[c]{ccccccccccccccccccc} 
 &  &  & \multicolumn{8}{c}{YH$_{6}$} &  &  & \multicolumn{6}{c}{YD$_{6}$}\tabularnewline
 \noalign{\vskip0.1cm}
 \hline 
 \noalign{\vskip0.2cm}
P &  &  & $v_{F}$ & $H_{c}\left(0\right)$ & $H_{c1}\left(0\right)$ & $H_{c2}\left(0\right)$ & $\kappa_{1}\left(0\right)$ & $\lambda_{L}\left(0\right)$ & $\xi\left(0\right)$ & &  & $v_{F}$  & $H_{c}\left(0\right)$ & $H_{c1}\left(0\right)$ & $H_{c2}\left(0\right)$ & $\kappa_{1}\left(0\right)$ & $\lambda_{L}\left(0\right)$ & $\xi\left(0\right)$\\
\noalign{\vskip0.2cm}
\hline 
160 &  &  & 8.76 & 3.79 & 0.33 & 141.5 & 26.37 & 159.15 & 1.712 & &  & 8.75 & 2.96 & 0.26 & 110.8 & 26.48 & 158.38 & 2.18\tabularnewline
\noalign{\vskip\doublerulesep}
180 &  &  & 8.68 & 3.59 & 0.33 & 129.0 & 25.57 & 157.98 & 1.884 & &  & 8.68 & 2.83 & 0.25 & 103.0 & 25.76 & 157.38 & 2.34\tabularnewline
\noalign{\vskip\doublerulesep}
200 &  &  & 8.61 & 3.45 & 0.31 & 122.3 & 25.04 & 157.40 & 1.954 & &  & 8.59 & 2.74 & 0.25 & 97.7 & 25.27 & 157.22 & 2.43\tabularnewline
\noalign{\vskip\doublerulesep}
220 &  &  & 8.53 & 3.35 & 0.31 & 116.7 & 25.64 & 158.12 & 2.164 & &  & 8.51 & 2.68 & 0.25 & 91.9 & 24.93 & 157.93 & 2.54\tabularnewline
\noalign{\vskip\doublerulesep}
250 &  &  & 8.39 & 3.25 & 0.30 & 111.6 & 24.27 & 159.04 & 2.110 & &  & 8.38 & 2.61 & 0.24 & 90.9 & 24.60 & 159.28 & 2.58\tabularnewline[\doublerulesep]
\end{tabular}
\end{ruledtabular}
\end{table*}

To summarize, the superconducting properties and electromagnetic field response of the yttrium hydride YH$_{6}$ are computed here from first principles, using DFT in conjunction with ME theory. From these, the experimental behavior of $T_{c}$ as a function of applied pressure and the isotopic effect can be perfectly reproduced within the harmonic approximation. Similarly, the calculated $H_{c2}(T)$ with the ME formalism shows excellent agreement with the available experimental data at temperatures near $T_{c}$. As $T$ goes to zero, it shows intermediate values between the GL and WHH models, providing an important strong-coupling correction to these currently used phenomenological models. The implemented formalism even allows a  full description of the $H-T$ phase diagram by calculating $H_{c}(T)$, $H_{c1}(T)$, and $\kappa_{1}(T)$. Finally, the description of the YH(D)$_6$ superconducting state is completed by calculating $\xi(T)$ and $\lambda_{L}(T)$ within the clean limit, as well as the energy gap $\Delta_{0}$. From our results, we found that any deviation from BCS behavior is well explained as a strong-coupling correction, and in this basis we are able to discard any possible anomalous behavior or departure from conventional superconductivity as was previously suggested ~\cite{adma.202006832}. Even more, we consider that ME theory, in conjunction with Rainer-Bergmann~\cite{Rainer1974} and Nam's~\cite{PhysRev.156.470,PhysRev.156.487} formalism, can provide a general prescription to describe the superconducting state in the high-$T_{c}$ metal hydrides.


\begin{acknowledgments}
This research is funded in part by the Gordon and Betty Moore Foundation's EPiQS Initiative, Grant GBMF10731,  partially supported by the Consejo Nacional de Ciencia y Tecnología (CONACyT, México) under Grant No. FOP16-2021-01-320399, as well as by the U.S. Department of Energy, Office of Basic Energy Sciences under Award Number DE-SC0020303. The authors thankfully acknowledge computer resources, technical advice, and support provided by Laboratorio Nacional de Superc{\'o}mputo del Sureste de M{\'e}xico (LNS), a member of the CONACYT national laboratories.
\end{acknowledgments}

\appendix

\bibliographystyle{unsrt}
\bibliography{YH6_fields}

\end{document}


\title{SUPPLEMENTAL MATERIAL \\
The Isotope Effect and Critical Magnetic Fields of Superconducting YH$_{6}$: A Migdal-Eliashberg Theory Approach}

\author{S. Villa-Cort{\'e}s}
\email{svilla@ifuap.buap.mx, sergio.cortes@unlv.edu}
\affiliation{\Puebla}
\affiliation{\NEXCL}
\author{O. De la Pe{\~n}a-Seaman}
\affiliation{\Puebla}
\author{Keith V. Lawler}
\affiliation{\NEXCL}
\author{Ashkan Salamat}
\affiliation{\NEXCL}
\affiliation{\LV}

\date{\today}





\maketitle

\section{Technical Details:}
\subsection{Details on ab initio calculations}

The calculations of the structural, electronic, lattice dynamics and electron-phonon (el-ph) coupling properties were carried out within the framework of Density Functional Theory (DFT) \citep{PhysRev.140.A1133} and Density Functional Perturbation Theory (DFPT) \citep{0953-8984-21-39-395502,RevModPhys.73.515}, both implemented in the QUANTUM ESPRESSO suit code \citep{0953-8984-21-39-395502}. The calculations for the electronic properties were performed with a $24\times24\times24$ $k$-point mesh, and a $60$~Ry cutoff for the plane-wave basis, while the Perdew-Burke-Ernzerhof (PBE) functional \citep{PBE}, using norm-conserving Vanderbilt pseudopotentials \cite{NCV} from the PseudoDojo Library~\cite{DJ}, was employed to take into account the exchange and correlation contributions. The matrix elements of the electron-phonon interaction were calculated over a denser $48\times48\times48$ $k$-point mesh.

Fourier's interpolation of dynamical matrices, calculated on a $8\times8\times8$ $q$-point mesh, were used to determine the complete phonon spectra of the studied systems. Corrections due to quantum fluctuations at zero temperature, zero-point energy (ZPE) effects, are estimated through the quasi-harmonic approximation (QHA)~\citep{10.2138/rmg.2010.71.3,PhysRevB.99.214504}. Thus, the electronic structure, lattice dynamics, and el-ph properties, calculated for the $lm\bar{3}m$ crystal structure at different applied pressure values, incorporate the ZPE correction.

\subsection{Ground state, lattice dynamics, and coupling properties}

\begin{figure}
\includegraphics[width=8.4cm]{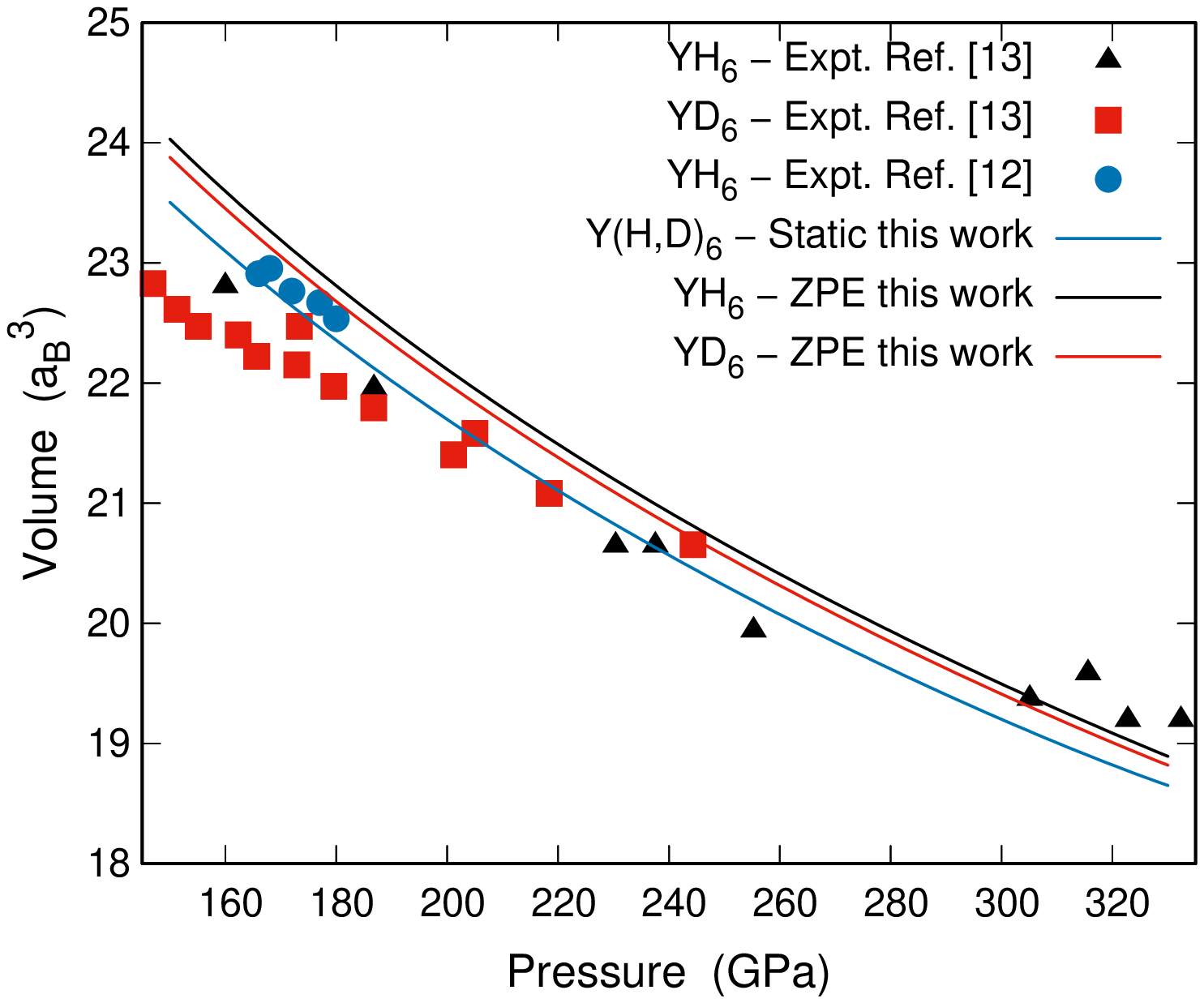}
\caption{\label{fig:PV}Calculated equation of state for $\textrm{Y}\textrm{H}_{6}$ and $\textrm{Y}\textrm{D}_{6}$ with (ZPE) and without (static) zero-point energy contribution. Experimental measurements from Troyan et al. \cite{adma.202006832} and  Panpan Kong et al. \cite{Kong2021} are also shown.}
\end{figure}

\begin{figure}
\includegraphics[width=8.4cm]{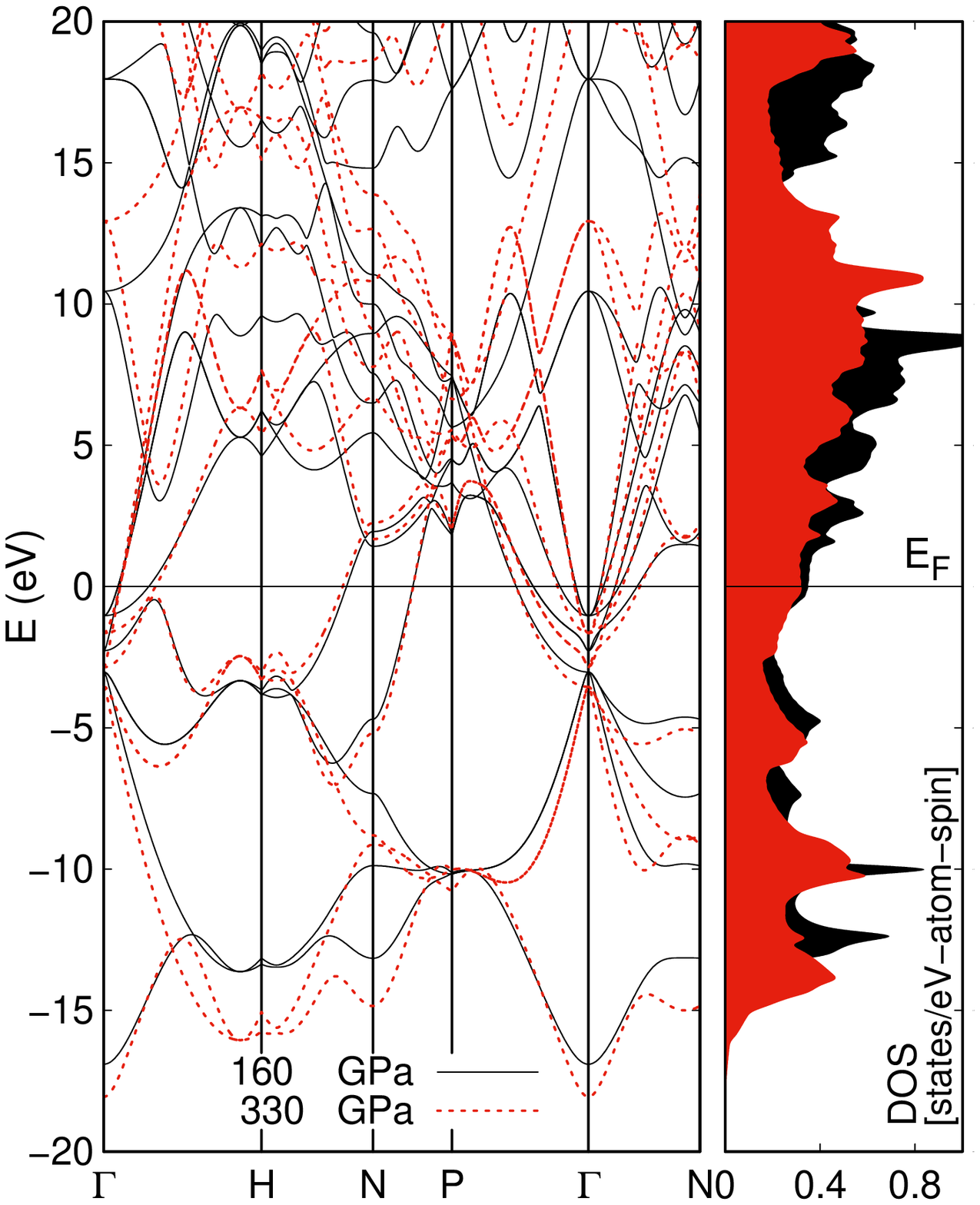}\caption{\label{fig:bands}Calculated electronic band structure and electronic density of states for YH$_{6}$ at 160 and 330 GPa.}
\end{figure}

\begin{figure*}
\includegraphics[width=17cm]{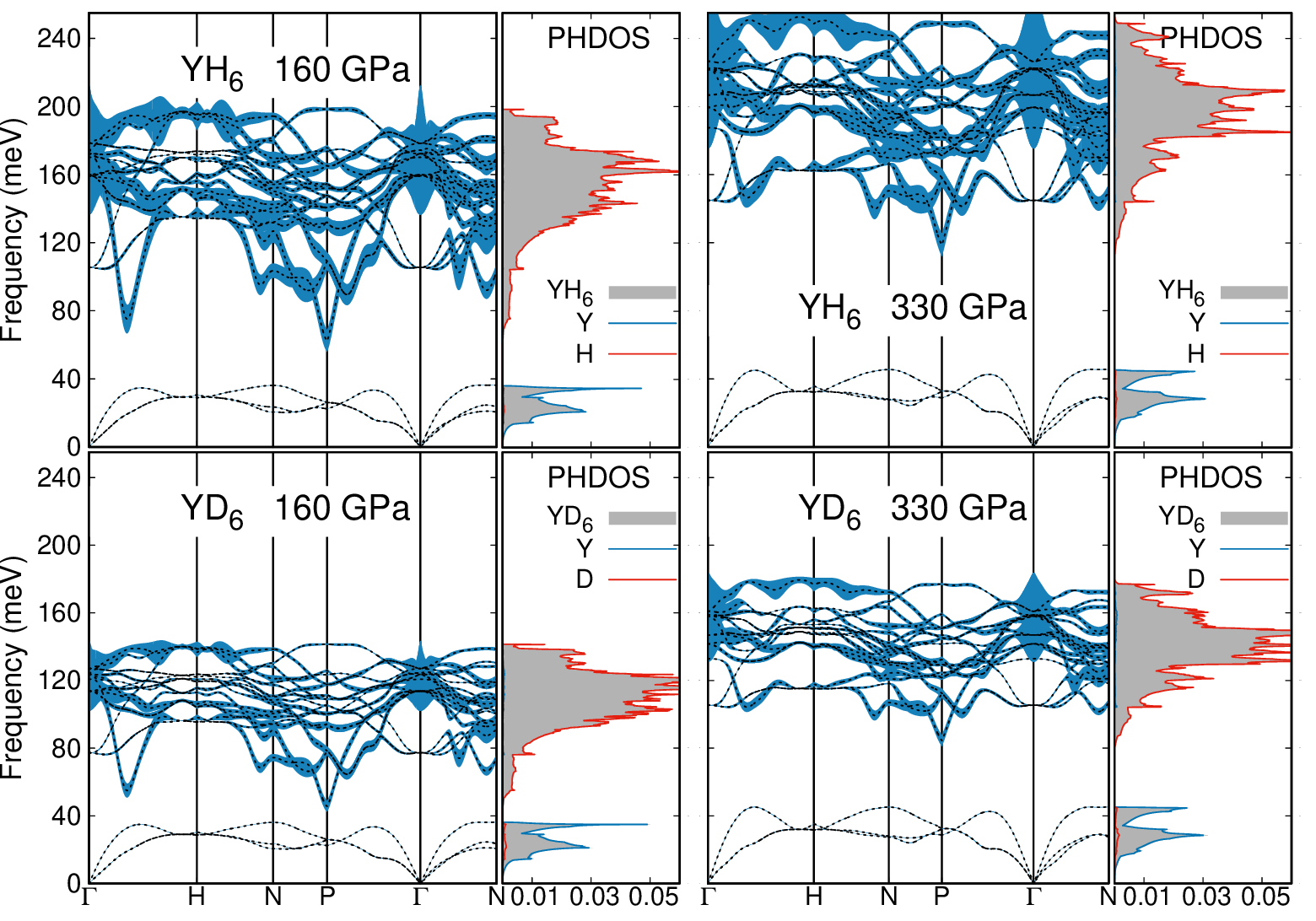}\caption{\label{fig:modos} Phonon dispersion and phonon density of states (total and projected) calculated for YH$_{6}$ (top) and YD$_{6}$ (bottom) at 160 and 330 GPa. The vertical lines at $\omega_{\vec{q}}$ represent the associated phonon linewidhts.}
\end{figure*}

\begin{figure*}
\includegraphics[width=17cm]{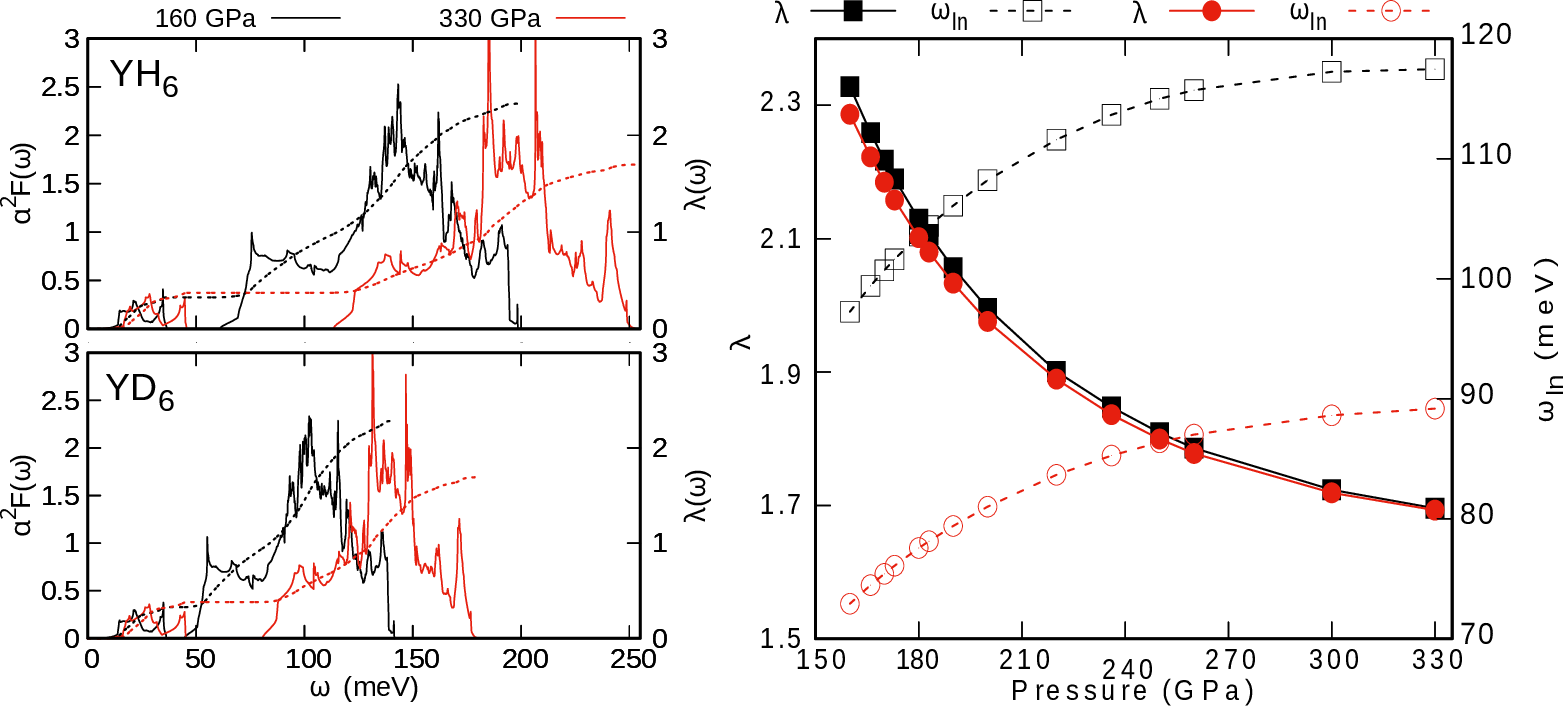}
\caption{\label{fig:lamb-wln}Eliashberg function (solid lines) and electron-phonon coupling constant, as a function of frequency, (dashed lines) for YH$_{6}$ and YD$_{6}$ at 160 and 330 GPa (a). Electron-phonon coupling constant ($\lambda$), and the Allen-Dynes characteristic phonon frequency ($\omega_{ln}$), for YH$_{6}$ (squares) and YD$_{6}$ (circles) as a function of the applied pressure (b).
}
\end{figure*}

In Fig. \ref{fig:PV} we present the equation of state for YH$_{6}$ and YD$_{6}$ with (ZPE) and without (static) considering the zero-point motion contribution to the energy. The crystal lattice parameters were optimized within QHA at pressures (from 160 to 330 GPa) where experimental values of $T_{c}$ have been measured. In Fig. \ref{fig:bands} we show the electronic band structure and the density of states (which include ZPE corrections) for YH$_{6}$ at some pressures of interest. In Fig. \ref{fig:modos} we show the phonon spectrum and the phonon density of states (PHDOS) for both compounds, YH$_{6}$ and YD$_{6}$, at 160 and 330 GPa, including ZPE. To gain more insight, Fig. \ref{fig:modos} also shows phonon linewidths of the $\vec{q}\nu$ phonon mode, $\gamma_{\vec{q}\nu}$, arising from the electron-phonon interaction given by~\cite{PhysRevB.6.2577,PhysRevB.9.4733}
\begin{equation}
\gamma_{\vec{q}\nu}=2\pi\omega_{\vec{q}\nu}\sum_{\vec{k},nm}\left|g_{\vec{k}+\vec{q},m,\vec{k},n}^{\vec{q}\nu}\right|^{2}\delta\left(\varepsilon_{\vec{k}+\vec{q},m}-\varepsilon_{F}\right)\delta\left(\varepsilon_{\vec{k},n}-\varepsilon_{F}\right),
\end{equation}
where $g_{\vec{k}+\vec{q},m,\vec{k},n}^{\vec{q}\nu}$ are the matrix elements of the electron-phonon interaction, $\varepsilon_{\vec{k}+\vec{q},m}$ and $\varepsilon_{\vec{k},n}$ are one-electron band energies, with band index $m$ and $n$, with vectors $\vec{k}+\vec{q}$ and $\vec{k}$, respectively, and $\omega_{\vec{q}\nu}$ is the phonon frequency for mode $\nu$ at wave-vector $\vec{q}$.
In Fig. \ref{fig:lamb-wln}(a) we show the Eliashberg function as well as the electron-phonon coupling parameter, $\lambda(\omega)$, for both compounds at 160 and 330 GPa. The Eliashberg function is defined as
\begin{equation}
\alpha^{2}F\left(\omega\right)= \frac{1}{N\left(\epsilon_{F}\right)}\sum_{\vec{q}\nu}\delta\left(\omega-\omega_{\vec{q}\nu}\right)\sum_{\vec{k},nm}\left|g_{\vec{k}+\vec{q},m,\vec{k},n}^{\vec{q}\nu}\right|^{2}\delta\left(\epsilon_{\vec{k}+\vec{q},m}-\epsilon_{F}\right)\delta\left(\epsilon_{\vec{k},n}-\epsilon_{F}\right),
\label{eq:a2f-def0}
\end{equation}
and the electron-phonon coupling parameter is given by
\begin{equation}
\lambda=2\int_{0}^{\infty}d\omega\frac{\alpha^{2}F\left(\omega\right)}{\omega},\label{eq:el-ph-parameter0}
\end{equation}
Finally, in Fig. \ref{fig:lamb-wln}(b) the evolution of $\lambda$, for YH$_{6}$ and YD$_{6}$, as pressure increases is shown, as well as the Allen-Dynes characteristic phonon frequency~\cite{PhysRevB.12.905} defined by 
\begin{equation}
\omega_{ln}=\exp\left\{ \frac{2}{\lambda}\int_{0}^{\infty}d\omega\frac{\alpha^{2}F\left(\omega\right)}{\omega}\ln\omega\right\} .
\end{equation}

%
%

\subsection{Details on Migdal-Eliashberg calculations}

For an isotropic superconductor in absence of a magnetic field, the Nonlinear Migdal-Eliashberg equations (NLMEE) are~\cite{Rainer1974,Daams1979}
\begin{eqnarray}
\tilde{\Delta}_{n} & = & \pi k_{B}T\sum_{m}\left(\lambda_{nm}-\mu^{*}(\omega_{c})\right)\frac{\tilde{\Delta}_{m}}{\left(\tilde{\omega}^2_{m}+\tilde{\Delta}^2_{m}\right)^{1/2}},\label{NLMEE1}
\end{eqnarray}
\begin{eqnarray}
\tilde{\omega}_{n} & = & \omega_{n}+\pi k_{B}T\sum_{m}\lambda_{nm}\frac{\tilde{\omega}_{m}}{\left(\tilde{\omega}^2_{m}+\tilde{\Delta}^2_{m}\right)^{1/2}},
\label{NLMEE2}
\end{eqnarray}
where $T$ is the temperature and $\omega_{n}$ are the Fermion Matsubara frequencies, expressed as $i\omega_{n}=i\pi k_{B}T\left(2n-1\right)$. Under this framework, the coupling parameter $\lambda_{nm}$ is defined as
\begin{equation}
\lambda_{nm}=2\int_{0}^{\infty}d\omega\frac{\omega\alpha^{2}F\left(\omega\right)}{\omega^{2}+\left(\omega_{m}-\omega_{n}\right)^{2}},\label{eq:el-ph-parameter}
\end{equation}
$\lambda_{nn}$ is the known electron-phonon coupling parameter ($\lambda$) and $\mu^{*}$ is the screened electron-electron repulsion parameter.%

Introducing $\bar{\Delta}_{n}=\tilde{\Delta}_{n}/\left[\left|\tilde{\omega}_{n}\right| + \rho\right]$, where $\rho$ is the breaking parameter that becomes zero at $T_{c}$, we get the Linearized Migdal-Eliashberg equations (LMEE)~\cite{Bergmann1973,VILLACORTES2022110451},
\begin{eqnarray}
\rho\bar{\Delta}_{n} & = & \pi k_{B}T\sum_{m}\left[\lambda_{nm}-\mu^{*}(\omega_{c})-\delta_{nm}\frac{\left|\tilde{\omega}_{n}\right|}{\pi T}\right]\bar{\Delta}_{m},
\label{LMEE1}
\end{eqnarray}
\begin{eqnarray}
\tilde{\omega}_{n} & = & \omega_{n}+\pi k_{B}T\sum_{m}\lambda_{nm}sgn(\omega_{m}),
\label{LMEE2}
\end{eqnarray}
Finally, from Eq. (\ref{LMEE1}) the functional derivative of $T_{c}$ respect to the Eliashberg function and $\mu^{*}$ are given by~\cite{Bergmann1973,VILLACORTES2022110451,Daams1979}
\begin{equation}
\frac{\delta T_{c}}{\delta\alpha^{2}F\left(\omega\right)}=-\left.\left(\frac{d\rho}{dT}\right)\right|_{T_{c}}\frac{\delta\rho}{\delta\alpha^{2}F\left(\omega\right)},\label{eq:dev_a2f}
\end{equation}
and the partial derivative of $T_{c}$ respect to $\mu^{*}$ is~\cite{Bergmann1973,VILLACORTES2022110451,Daams1979} 
\begin{equation}
\frac{\partial T_{c}}{\partial\mu^{*}}=-\left.\left(\frac{d\rho}{dT}\right)\right|_{T_{c}}\frac{\partial\rho}{\partial\mu^{*}}.\label{eq:dev_mu}
\end{equation}

To study the thermodynamics at temperatures below the critical temperature for each pressure, we use the solutions of the NLMEE (Eq. \ref{NLMEE1}), from which the free energy difference $\Delta F(T)$ between the superconducting and normal states can be computed. It is given by~\cite{Rainer1974}
\begin{eqnarray}
\Delta F(T)= &  & N(0)\pi T\sum_{n}\left[2\left|\tilde{\omega}_{n}\right|\left(\frac{1}{\sqrt{1+\bar{\Delta}_{n}^{2}}}-1\right)\right.\nonumber \\
 &  & +\pi T\sum_{m}\left(\lambda_{nm}-\mu^{*}\right)\frac{\bar{\Delta}_{n}}{\sqrt{1+\bar{\Delta}_{n}^{2}}}\frac{\bar{\Delta}_{m}}{\sqrt{1+\bar{\Delta}_{m}^{2}}}\\
 &  & \left.-\pi T\sum_{m}\left(1-\frac{1}{\sqrt{1+\bar{\Delta}_{n}^{2}}}\frac{1}{\sqrt{1+\bar{\Delta}_{m}^{2}}}\right)\lambda_{nm}sgn\left(\omega_{m}\omega_{n}\right)\right]\nonumber 
\end{eqnarray}
 where $N(0)$ is the single-spin density of electronic states at the Fermi energy. \\
 
 The upper critical field, $H_{c2}(T)$, has to be evaluated from the LMEE valid in the presence of a homogeneous magnetic field~\cite{PhysRev.158.332,PhysRev.158.415,PhysRevB.2.135,Rainer1974}. For a dirty strong-coupling superconductor these equations are
\begin{eqnarray}
\tilde{\Delta}_{n} & = & \pi k_{B}T\sum_{m}\left(\lambda_{nm}-\mu^{*}(\omega_{c})\right)\frac{\tilde{\Delta}_{m}}{\left|\tilde{\omega}_{n}\right| + \rho\left(T\right)},
\label{NLMEEHC2}
\end{eqnarray}
\begin{eqnarray}
\tilde{\omega}_{n} & = & \omega_{n}+\pi k_{B}T\sum_{m}\lambda_{nm}sgn(\omega_{m}).
\label{NLMEE2HC2}
\end{eqnarray}
Which resemble Eqs. (\ref{LMEE1}) and (\ref{LMEE2}), with the difference that in those Eqs. the pair-breaking parameter $\rho$ is introduced as a mathematical artifact that holds only at $T_{c}$ and is used to solve the NLMEE. Here the pair-breaking parameter $\rho\left(T\right)$ is determined by the condition that Eq. (\ref{NLMEEHC2}) has a nontrivial solution $\tilde{\Delta}_{n}$, and is related to $H_{c2}(T)$ by
\begin{equation}
H_{c2}(T)=\rho(T)/eD,\label{eq:FHC2}
\end{equation}
\begin{equation}
D=l_{tr}v_{F}/3,\label{eq:DHC2}
\end{equation}
 where $D$ is the diffusion constant, $v_{F}$ the Fermi velocity, $l_{tr}$ the mean free path, and $e$ is the electron charge.
\subsection*{T$_{c}$ under pressure}

The considered changes are between two different applied pressures, $P_i$ and $P_{i+1}$. Then, for the Eliashberg function it's change would be
$\Delta \alpha^{2}F\left(\omega,P_{i},P_{i+1}\right) = \alpha^{2}F\left(\omega,P_{i+1}\right) - \alpha^{2}F\left(\omega,P_{i}\right)$, where both, $\alpha^{2}F\left(\omega,P_{i+1}\right)$ and $\alpha^{2}F\left(\omega,P_{i}\right)$ are calculated independently. The resulting change $\Delta T_c$ of the critical temperature from the system at $P_i$ to the system at $P_{i+1}$ is given by~\cite{Bergmann1973,VILLACORTES2018371,VILLACORTES2022110451}
\begin{eqnarray}
\Delta T_{c}(P_i) & = & \int_{0}^{\infty}\frac{\delta T_{c}(P_i)}{\delta\alpha^{2}F\left(\omega\right)}\Delta\alpha^{2}F\left(\omega,P_i,P_{i+1}\right)d\omega,\label{Dtc}
\end{eqnarray}
and the critical temperature for the system under pressure $P_{i+1}$ is then 
\begin{eqnarray}
T_{c}(P_{i+1})=T_{c}(P_{i})+\Delta T_{c}(P_{i}).\label{tp}
\end{eqnarray}
\subsection*{The isotope effect}
\begin{figure}
\includegraphics[width=8.4cm]{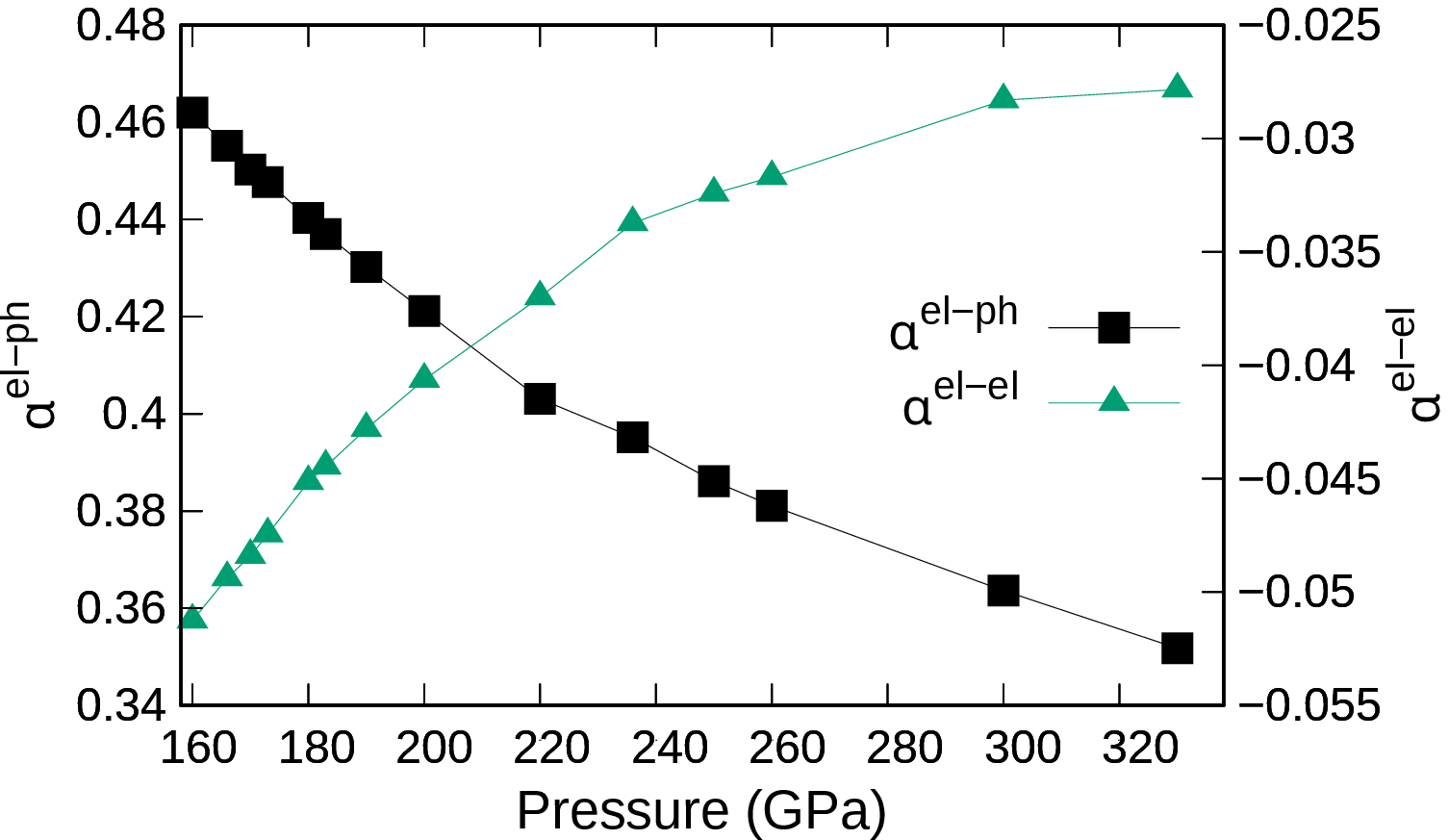}\caption{\label{fig:iso}Calculated electron-phonon and electron-electron contributions to the isotope effect as a function of pressure for YH$_{6}$.}
\end{figure}
According to Migdal-Eliashberg theory, $T_{c}$ depends on both, the electron-phonon interaction and the electron-electron repulsion. Then, taking that into account, the partial isotope coefficients can be split into two contributions~\cite{meromio,VILLACORTES2018371,VILLACORTES2022110451}: one that comes from the change in $T_{c}$ due to changes in the electron-phonon interaction, namely $\alpha^{el-ph}$, while the second one comes from the change in $T_{c}$ due to changes in the electron-electron interaction, $\alpha^{el-el}$. The partial isotope coefficient is then given by 
\begin{equation}
\alpha_{i}=\alpha^{el-ph}+\alpha^{el-el}.\label{eq:Totalalfa}
\end{equation}
Here, $\alpha^{el-el}$ is defined as
\begin{equation}
\alpha^{el-el}=-\frac{\Delta T_{c}^{el-el}}{T_{c}\Delta\ln M},
\label{alphaelel}
\end{equation}
where the change in $T_{c}$ due to the electron-electron interaction is expressed as~\cite{meromio}
\begin{equation}
\Delta T_{c}^{el-el}=\frac{\partial T_{c}}{\partial\mu^{*}}\left(\mu_{YD_{6}}^{*}-\mu_{YH_{6}}^{*}\right).\label{eq:tcel-el}
\end{equation}
In a similar way, $\alpha^{el-ph}$ is defined as
\begin{equation}
\alpha^{el-ph}=-\frac{\Delta T_{c}^{el-ph}}{T_{c}\Delta\ln M},
\label{alphaelph}
\end{equation}
where the change in $T_{c}$ due to the electron-phonon interaction is calculated as
\begin{eqnarray}
\Delta T_{c}^{el-ph} & = & \int_{0}^{\infty}\frac{\delta T_{c}}{\delta\alpha^{2}F\left(\omega\right)}\Delta\alpha^{2}F\left(\omega,H,D\right)d\omega,\label{DtcIso}
\end{eqnarray}
with $\Delta\alpha^{2}F\left(\omega,H,D\right)$ the difference in the calculated Eliashberg functions for YH$_{6}$ and YD$_{6}$. In Figure \ref{fig:iso} we show the electron-phonon and electron-electron contributions to the isotope effect calculated using Eqs. (\ref{alphaelph}) and (\ref{alphaelel}) respectively. \\
\subsection*{Numerical aspects}

To solve the LMEE and NLMEE, a cut-off frequency $\omega_{c}$ is defined in terms of the maximum phonon frequency $\omega_{ph}$, in order to cut the sum over the Matsubara frequencies. We perform our calculations with a cut-off of $\omega_{c}=600\times\omega_{ph}$, which gives us a minimum of 1500 Matsubara frequencies and ensures good numerical convergence. Once $\omega_{c}$ is set, the only adjustable parameter left is the cut-off dependent $\mu^{*}(\omega_{c})$, which is related to the cut-off independent $\mu^{*}$ through the relation~\cite{PhysRevB.12.905}
\begin{equation}
\frac{1}{\mu^{*}(\omega_{c})}=\frac{1}{\mu^{*}}+ln\left(\frac{\bar{\omega}_{2}}{\omega_{c}}\right),
\end{equation}
with
\begin{equation}
\bar\omega_{2}=\left(\frac{2}{\lambda}\int d\omega \alpha^{2}F(\omega)\omega\right)^{1/2}.
\end{equation}
It is $\mu^{*}$ (rather than $\mu^{*}(\omega_{c})$) which has to become regarded as physically significant, and which is frequently deduced from tunneling experiments or isotope-effect measurements~\cite{PhysRevB.12.905}. Here, $\mu^{*}$ was fitted to get the experimental $T_{c}$ at each pressure of interest. Our fitted $\mu^{*}$ was $0.129$ for YH$_{6}$ with $T_{c}=$220 K at 160 GPa, and $0.127$ for YD$_{6}$ with $T_{c}=$165 K at 200 GPa. Once Eq. (\ref{tp}) is used to get $T_{c}(P)$ at every other pressure, $\mu^{*}$ is fitted at each new $T_{c}(P)$. \\

The average Fermi velocity in Eq. (\ref{eq:DHC2}) was calculated from DFT band structure using,
\begin{equation}
\left<v_{F}\right>=\left[\frac{\sum_{k}\delta(E_{k}-E_{F})v^{2}}{\sum_{k}\delta(E_{k}-E_{F})}\right]^{1/2}.
\end{equation}

To calculate $v_{F}$, a very dense $72\times72\times72$ $k$-point mesh was used. The mean free path ($l_{tr}$) was fitted using Eqs. (\ref{eq:DHC2}) and (\ref{eq:FHC2}) to get the experimental $H_{c2}(T)$ values near $T_{c}$. For YH$_{6}$ at 160 and 200 GPa we found $l_{tr}$ to be 1.655 \AA  and for DH$_{6}$ at 173 GPa the fitted value was 1.645 \AA. \\

\begin{figure}
\includegraphics[width=8.4cm]{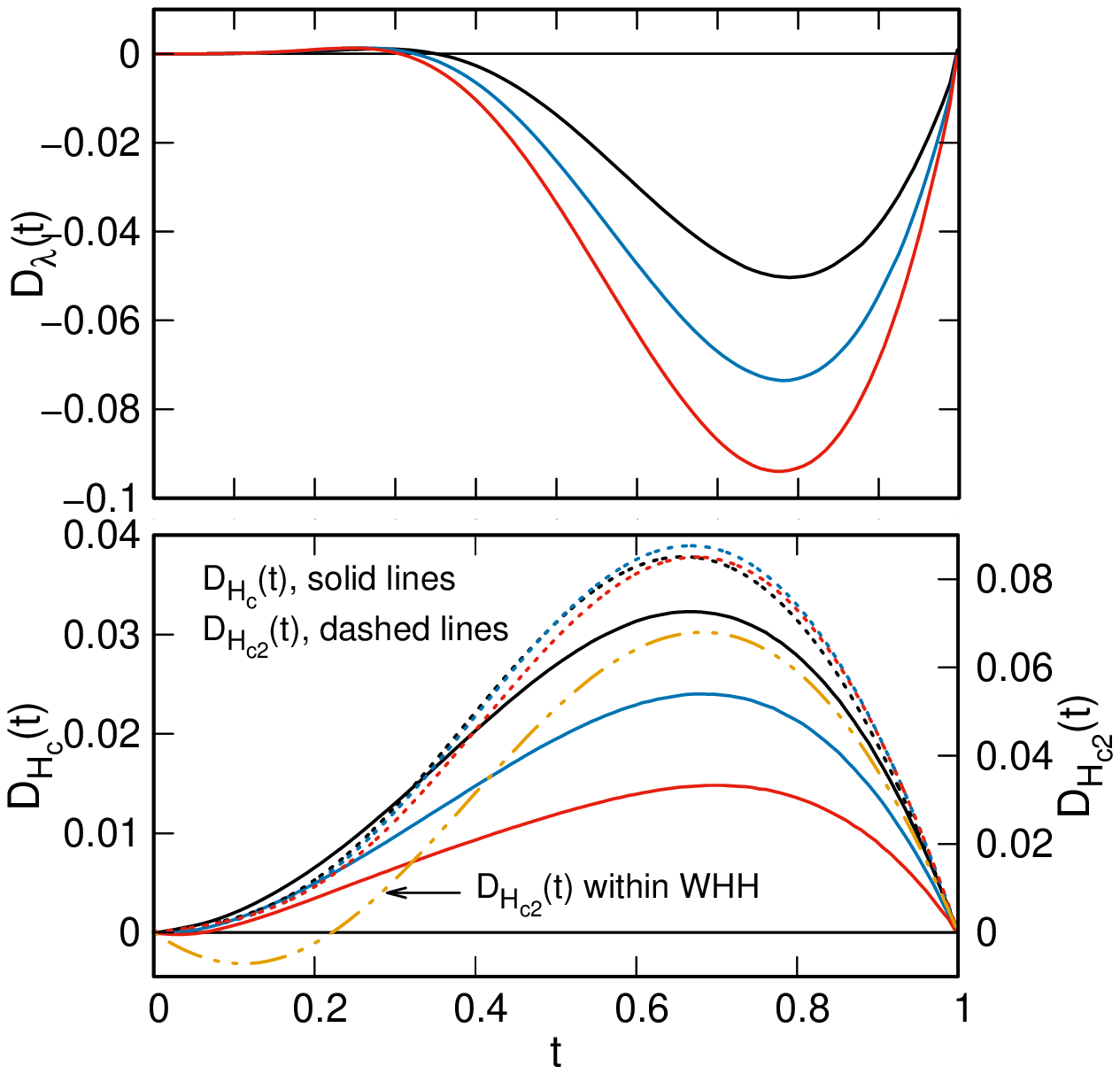}\caption{\label{fig:deviation}Deviation function for YH$_{6}$ at 160 (black), 200 (blue), and 250 (red) GPa for the thermodynamic critical field $H_{c}(T)$, the upper critical field $H_{c2}$(T), and the London penetration depth $\lambda_{L}(T)$.}
\end{figure}

Finally, Figure \ref{fig:deviation} shows the deviation functions for $H_{c}$, $H_{c2}$, and $\lambda_{L}$ defined as
\begin{equation}
D_{H_{c}}(t)=\frac{H_{c}(t)}{H_{c}(0)} - (1-t^{2}),
\end{equation}
\begin{equation}
D_{H_{c2}}(t)=\frac{H_{c}(t)}{H_{c}(0)} - \frac{1-t^{2}}{1+t^{2}},
\end{equation}
\begin{equation}
D_{\lambda_{L}}(t)=\frac{\lambda_{L}^{2}(0)}{\lambda_{L}^{2}(t)} - 1-t^{4},
\end{equation}
with $t=T/T_{c}$.


\bibliography{YH6_sm}